\documentclass[conference, a4paper]{IEEEtran}

\usepackage{algorithm,algorithmic}
\usepackage{amsmath,amssymb,mathrsfs}
\usepackage{cite}
\usepackage{subfigure}
\usepackage{graphicx,color}
\usepackage[nomain,acronym,toc]{glossaries}
\usepackage{subfigure}
\usepackage{balance}
\usepackage{tikz,siunitx}
\usepackage{url}

\newtheorem{cor}{Corollary}

\newtheorem{prop}{Proposition}

\newacronym{2d}{2-D}{two-dimensional}
\newacronym{3d}{3-D}{three-dimensional}
\newacronym{3gpp}{3GPP}{3rd Generation Partnership Project}
\newacronym{ap}{AP}{access point}
\newacronym{awgn}{AWGN}{additive white Gaussian noise}
\newacronym{ber}{BER}{bit-error-rate}
\newacronym{bs}{BS}{base station}
\newacronym{cbsm}{CBSM}{correlation-based stochastic models }
\newacronym{ccdf}{CCDF}{complementary cumulative distribution function}
\newacronym{cdf}{CDF}{cumulative distribution function}
\newacronym{csit}{CSIT}{channel-state-information at the transmitter}
\newacronym{dof}{DoF}{degree of freedom}
\newacronym{elaa}{ELAA}{extremely large aperture array}
\newacronym{fdma}{FDMA}{frequency-division multiple-access}
\newacronym{gbsm}{GBSM}{geometry-based stochastic models}
\newacronym{harq}{HARQ}{hybrid automatic repeat request}
\newacronym{iid}{i.i.d.}{independent and identically distributed}
\newacronym{ind}{i.n.d.}{independent and non-identically distributed}
\newacronym{inf}{InF}{Indoor Factory}
\newacronym{itu-r}{ITU-R}{International Telecommunication Union Radiocommunication Sector}
\newacronym{los}{LoS}{line-of-sight}
\newacronym{lmmse}{LMMSE}{linear minimum mean square error}
\newacronym{mac}{MAC}{medium-access control}
\newacronym{mmse}{MMSE}{minimum mean square error}
\newacronym{ms}{MS}{mobile station}
\newacronym{nlos}{NLoS}{non-line-of-sight}
\newacronym{oamp}{OAMP}{orthogonal approximate message passing}
\newacronym{ofdma}{OFDMA}{orthogonal frequency-division multiple-access}
\newacronym{pdf}{PDF}{probability density function}
\newacronym{pmf}{PMF}{probability mass function}
\newacronym{pwm}{PWM}{plane wave model}
\newacronym{rss}{RSS}{received signal strength}
\newacronym{se}{SE}{spectral efficiency}
\newacronym{snr}{SNR}{signal-to-noise ratio}
\newacronym{tti}{TTI}{transmission time interval}
\newacronym{ue}{UE}{user equipment}
\newacronym{ula}{ULA}{uniform linear array}
\newacronym{uma}{UMa}{urban macro}
\newacronym{umi}{UMi}{urban micro}
\newacronym{urllc}{URLLC}{ultra-reliable low-latency communication}
\newacronym{vr}{VR}{visibility region}
\newacronym{wss}{WSS}{wide-sense stationary}

\newglossaryentry{elamimo}
{
	name={ELAA-mMIMO},
	description={anyway},
	first={ELAA-mMIMO},	
	firstplural={ELAA-mMIMO},
	text={ELAA-mMIMO},
	plural={ELAA-mMIMO}	
}

\newglossaryentry{mmimo}
{
	name={mMIMO},
	description={anyway},
	first={mMIMO},	
	firstplural={mMIMO},
	text={mMIMO},
	plural={mMIMO}	
}

\definecolor{sblue}{RGB}{0,51,120}
\definecolor{sred}{RGB}{200,51,130}

\renewcommand{\eqref}[1]{(\ref{#1})}
\newcommand{\figref}[1]{Fig. \ref{#1}}

\ifCLASSINFOpdf
\else
\fi

\begin{document}

\title{A Non-Stationary Channel Model with Correlated NLoS/LoS States for ELAA-mMIMO}

\author{Jiuyu Liu$^{1}$, Yi Ma$^{1}$, Jinfei Wang$^{1}$, Na Yi$^{1}$, Songyan Xue$^{2}$, Rahim Tafazolli$^{1}$, and Fan Wang$^{2}$\\
	{\small $^{1}$5GIC and 6GIC, Institute for Communication Systems, University of Surrey, Guildford, UK, GU2 7XH}\\
		{\small $^{1}$Emails: (jiuyu.liu, y.ma, jinfei.wang, n.yi, r.tafazolli)@surrey.ac.uk}\\
		{\small $^{2}$Huawei Technologies Co., Ltd. Email: (xuesongyan, fan.wang)@huawei.com}
		}
\markboth{}%
{}

\maketitle

\begin{abstract}
In this paper, a novel spatially non-stationary channel model is proposed for link-level computer simulations of massive multiple-input multiple-output (mMIMO) with extremely large aperture array {(ELAA)}. 
The proposed channel model allows a mix of non-line-of-sight (NLoS) and LoS links between a user and service antennas. 
The NLoS/LoS state of each link is characterized by a binary random variable, which obeys a correlated Bernoulli distribution.
The correlation is described in {the} form of an exponentially decaying window. 
In addition, the proposed model incorporates shadowing effects which are non-identical for NLoS and LoS states.
It is demonstrated, through computer emulation, that the proposed model can capture almost all spatially non-stationary fading behaviors of the {ELAA-mMIMO} channel.
Moreover, it has a low implementational complexity. 
With the proposed channel model, Monte-Carlo simulations are carried out to evaluate the channel capacity of {ELAA-mMIMO}. 
{It is shown that the ELAA-mMIMO} channel capacity has considerably different stochastic characteristics from the conventional mMIMO due to the presence of channel spatial non-stationarity. 
\end{abstract}

\section{Introduction}\label{sec1}
\begin{figure}[t]
	\centering
	\includegraphics[scale=0.13]{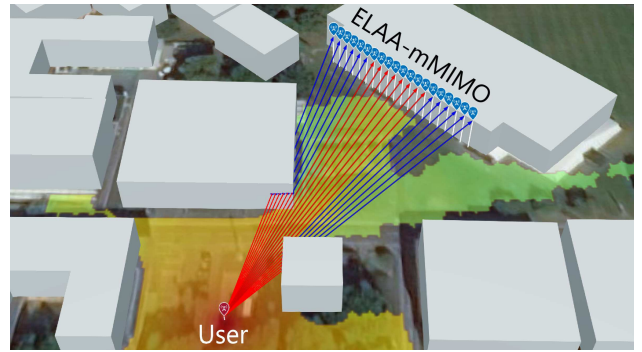}	
	\caption{A practical example of non-stationary ELAA-mMIMO channel with correlated NLoS/LoS conditions;
		{\color{red}$\longrightarrow$} LoS link, {\color{blue}$\longrightarrow$} NLoS link.}
	\label{fig01}
	\vspace{-2em}
\end{figure}

Massive multiple-input multiple-output (mMIMO) is a scalable multi-antenna signal transmission and reception technology due to the excess use of service antennas over user antennas. 
Assuming the mMIMO fading channel to be spatially wide-sense stationary uncorrelated scattering (WSSUS), the instantaneous mMIMO channel capacity shows decreasing fluctuations with the increase of service antennas. 
The fluctuations become negligibly small when the number of service antennas goes extremely large. 
This is so called the channel hardening effect in \cite{Hochwald2004}. 
However, the hypothesis of channel spatial stationarity is not always true, particularly for the emerging mMIMO technology using an extremely large aperture array (ELAA) \cite{BJORNSON20193}. For instance, let us consider an ELAA-mMIMO system with a thousand of co-located {uniform-linear-array (ULA)} antennas.  
When the system operates at the central frequency of $3.5$ GHz, the aperture of the antenna array could be typically $43$ meters or longer. 
In such circumstance, the Rayleigh distance could reach {tens thousand} of meters long, and users are located in the near-field of the ELAA \cite{8949454}.
Spherical wavefront instead of plane wavefront should be considered in the large-scale radio propagation model \cite{7414041}.
As illustrated in \figref{fig01}, some user-to-service antenna links are non-line-of-sight (NLoS), and the others are LoS. 
Shadowing effects can also vary from link to link \cite{6410305,7063445,7062910,6206345,6810277,9170651}. 
All of these physical characteristics render the ELAA-mMIMO channel spatially non-stationary.

There are mainly two types of channel models: {\em 1)} deterministic channel models, 
and {\em 2)} statistical channel models. The latter are more useful for link-level computer simulations since they are usually simple and tractable. 
Specifically for the ELAA-mMIMO non-stationary channel, deterministic models are surely too complex to implement. 
Statistical non-stationary channel models can be hardly made as simple as WSSUS channel models. 
A straightforward approach is to make extensions from WSSUS channel models by incorporating some spatial non-stationarity. 
For instance, independent and identically distributed (i.i.d.) Rayleigh (or Rician) fading channels can be extended to i.n.d. Rayleigh (or Rician) fading channels. 
The concept of visibility region is introduced to describe where in the ELAA user-to-service antenna links have their received signal strength (RSS) going beyond a threshold. 
The distribution of visibility region is dependent on the geometry of wireless environments. 
For instance in the literature \cite{8644126,8638522,9257469,9145378}, it is assumed that there exists only a single cluster between each user and service antennas. 
The length of visibility region obeys a log-normal distribution. When there exist multiple clusters, the distribution of visibility region for each cluster is modeled as the birth-death process or the Markov process. {The LoS states are based on the geometry of the wireless environment}  \cite{8125724,8866736,9120570,Amiri2020DistributedRF,bian2021general}. 
We found the single-cluster model a bit too simple to capture the channel spatial non-stationarity, {and on the other hand} the multi-cluster model is too complex {for} link-level computer simulations.   
This motivates us to develop a novel non-stationary statistical channel model for ELAA-mMIMO. 

In the proposed channel model, we fade out the concept of visibility region and allow user-to-service antenna links to be NLoS, LoS or a mix of them. 
The NLoS/LoS state of each link is described by a binary random variable, which obeys a correlated Bernoulli distribution. 
More specifically, the NLoS/LoS probability of each link is a function of the two-dimensional (2-D) distance between the user and the corresponding service antenna, which is defined by the \gls{3gpp} in \cite{3gpp.38.901}. 
Moreover, the probability for two arbitrary links to share the same NLoS/LoS state decreases exponentially with the separation between service-antennas. 
Therefore, the NLoS/LoS-state correlation is modeled as an exponentially decaying window. 
In addition, shadowing effects are carefully incorporated into the proposed non-stationary channel model. 
Within a NLoS/LoS window, all links share an identical NLoS/LoS state. 
They also share identical shadowing effects, which follow a log-normal distribution.
Both the NLoS/LoS state and shadowing effects vary from window to window. 
When a window experiences deep fades, this window can count as an invisible region and vice versa. 

In this paper, the realization of the proposed channel model is described in a pseudocode.
It has a simple structure and easy to implement in computer simulations. 
With the proposed channel model, we conduct Monte-Carlo simulations to evaluate the channel capacity of ELAA-mMIMO. 
It is observed that the channel spatial non-stationarity can be detrimental to the channel hardening effect in some cases. 
This could set a new and interesting task to the ELAA-mMIMO signal transmission and reception.

\section{System Model and Problem Statement}
ELAA-mMIMO has no fundamental differences from the conventional mMIMO in their mathematical forms. 
Taking the uplink communication as an example, their narrowband signal model can be both represented in the following standard vector/matrix form
\begin{equation}\label{eqn01}
	\mathbf{y = Hx + v},
\end{equation}
where $\mathbf{y} \in \mathbb{C}^{M\times 1}$ stands for the received signal block in the ELAA, 
$\mathbf{x} \in \mathbb{C}^{N \times 1}$ for the block of transmitted symbols \footnote{Each element of $\mathbf{x}$ represents a symbol sent by a transmit antenna. Symbols can be either from a single user or from multiple users.}, 
and $\mathbf{v} \sim \mathcal{CN}(0,\sigma_v^2\mathbf{I}_M)$ for the \gls{awgn}; $\mathbf{I}_M$ denotes an $(M)\times(M)$ identity matrix. 
For the conventional mMIMO system, it is often assumed: 

{\em A1)} The ratio $r\triangleq(M)/(N)$ must be sufficiently large; typically around $10$ or more.

{\em A2)} The number of service antennas $M$ must be very large (around $100 \sim 10,000$).
~\\
In addition to the above assumptions, ELAA has the following assumption:

{\em A3)} The service-antenna array must have an extremely large aperture. As already discussed in Section \ref{sec1}, it is in the length around $10$ or even $100$ meters. 

Main interests of this paper are in the random channel-transition matrix $\mathbf{H} \in \mathbb{C}^{M \times N}$. 
Providing the assumption {\em A3)}, elements of $\mathbf{H}$ do not obey an identical distribution, i.e., they can no longer be assumed as i.i.d. Rayleigh (or Rician, etc).
This is the channel spatial non-stationarity that must be appropriately modeled for computer simulations as well as analytical study.
Models must not be made too complex as they might not be simulation friendly and can be mathematically intractable.
On the other hand, they should not be made too simple to miss out some major physical characteristics. 
These requirements motivate the novel non-stationary channel model presented in Section \ref{sec3}.
\section{Novel Non-Stationary Channel Model}\label{sec3}
In the linear model \eqref{eqn01}, every element of $\mathbf{H}$ is corresponding to the channel coefficient of the link connecting a user-antenna to a service-antenna. 
To simplify mathematical notations, our discussion will be mainly focused on a single column of $\mathbf{H}$, denoted by $\mathbf{h}$. 
The $m$-th element of $\mathbf{h}$, denoted by $h_m, ~_{0\leq m\leq M-1}$, is the channel coefficient of the user-to-the $m$-th service-antenna link. 
Correlations between columns of $\mathbf{H}$ will be briefly discussed in Section \ref{sec04} and Section \ref{sec05}. 
With all the above in mind, the proposed channel model will be introduced step-by-step from the spherical-wave modeling, mixed NLoS/LoS states, 
channel normalization, and all the way through to the pseudocode for implementation.  

\subsection{i.n.d. Rayleigh or Rician Channel Models}
The extremely large aperture of ELAA-mMIMO requires the use of spherical-wave model to measure the Euclidean distance between the user and service antennas. In short, the Euclidean distance ($d_m$) of the $m$-th link varies with the link index $m$, and so as for the channel gain $|h_m|$. 
Taking this physical characteristic into account, the conventional i.i.d. Rayleigh channel model must be extended to the following i.n.d. Rayleigh channel model; as already used in \cite{8644126} for the link-level simulation
\begin{equation}\label{eqn02}
	h_{m} = h_m^{\textsc{nlos}}\triangleq\left(\frac{\alpha}{d_{m}^\rho}\right) \omega_m, ~_{m=0,..., M-1},
\end{equation}
where $\alpha, \rho$ are the path-loss attenuation and exponent, respectively, in the NLoS state, 
and $\omega_{m}$ is the complex Gaussian random variable often used for the realization of Rayleigh-fading channel.
Analogously, in the LoS state, the i.i.d. Rician channel should be extended to the following i.n.d. version
\begin{equation}\label{eqn03}	
	h_{m} = h_{m}^\textsc{los} \triangleq \frac{\beta}{d_{m}^{q}} \left(\sqrt{\dfrac{\kappa}{\kappa + 1}}\phi_{m} + \sqrt{\dfrac{1}{\kappa + 1}}\omega_{m} \right),
\end{equation}
where $\beta, q$ are the path-loss attenuation and exponent, respectively, in the LoS state,
$\phi_{m}\triangleq \exp(-j\frac{2\pi}{\lambda}d_{m})$ is the phase of the LoS path, 
and $\kappa$ is the Rician K-factor following a log-normal distribution, i.e., $\kappa\sim\mathcal{LN}(\mu_\kappa, \sigma_\kappa)$. 

\subsection{Integration of NLoS/LoS Non-Stationarity}
As already discussed in Section \ref{sec1} (also see \figref{fig01}),  an ELAA-mMIMO channel can involve NLoS links, LoS links or a mix of them. 
To incorporate this important characteristic, we propose a NLoS and LoS integrated channel model
\begin{equation}\label{eqn04}	
	h_{m} = b_{m}(\varepsilon_{m}^\textsc{los})^{\frac{1}{2}}h_{m}^\textsc{los} + (1-b_{m})(\varepsilon_{m}^\textsc{nlos})^{\frac{1}{2}}h_{m}^\textsc{nlos},
\end{equation}
where $\varepsilon_{m}^\textsc{los}\sim \mathcal{LN}(0,\sigma_\textsc{los}), \varepsilon_{m}^\textsc{nlos}\sim \mathcal{LN}(0,\sigma_\textsc{nlos})$ represent the shadowing effects in LoS or NLoS state, respectively. 
In \eqref{eqn04}, the most appealing factor is the binary random variable $b_m\in\{0, 1\}$. For the case of $b_m=0$, \eqref{eqn04} turns into the Rayleigh channel \eqref{eqn03} with NLoS shadowing effects; or otherwise for $b_m=1$, \eqref{eqn04} turns into the Rician channel \eqref{eqn02} with LoS 
shadowing effects. 
In practical wireless environments, $b_m$ cannot be independently generated with respect to the link index $m$.
Therefore, it is important to understand the probability distribution of $b_m, \forall m$.
\begin{prop}\label{prop1}
For $\ell\in\{0,...,M-1\}$, suppose that $b_\ell$ obeys the Bernoulli distribution $f(b_\ell; p)$, i.e.,
\begin{equation}\label{eqn05}
f(b_\ell; p)\triangleq p^{b_\ell}(1-p)^{1-b_\ell},~b_\ell\in\{0, 1\},
\end{equation}
where $p$ is the probability for $b_\ell=1$, and $(1-p)$ the probability for $b_\ell=0$.
The probability mass function (PMF) of $b_m, _{\forall m\neq \ell}$ is $f(b_m; p_m)$, and $p_m$ is given by
\begin{equation}\label{eqn06}
p_m=(2p-1)\exp\left(-\frac{\Delta(\ell,m)}{d_\mathrm{cor}}\right)+1-p,
\end{equation}
where $\Delta(\ell, m)$ is the distance between two service-antennas, and $d_\mathrm{cor}$ the correlation distance based on the propagation environment.
\end{prop}

{\em Proposition \ref{prop1}} gets well supported by measurement results in \gls{3gpp} technical documents. 
In \cite[(Table 7.4.2-1)]{3gpp.38.901}, the probability of LoS state is described as a function of the 2-D distance ($d^\mathrm{2D}$) between the service antenna and the user
\footnote{The definition of 2-D distance is specified inside the \gls{3gpp} document.}.
For instance in the \gls{umi} scenario, the probability for a link to be LoS is given by
\begin{IEEEeqnarray}{ll}\label{eqn07}
	\mathcal{P}_{\textsc{los}}(d^\textsc{2d}) = & \min\left(\frac{\overline{d_1}}{d^\textsc{2d}},1\right) \left[1-\exp\left(-\frac{d^\textsc{2d}}{\overline{d_2}}\right)\right] \nonumber\\
	&\quad\quad\quad\quad\quad\quad\quad\quad+ \exp \left(-\frac{d^\textsc{2d}}{\overline{d_2}}\right),
\end{IEEEeqnarray}
where $\overline{d_1}$, $\overline{d_2}$ are the reference distances that can be found in \cite{3gpp.38.901} for various scenarios.
For the $\ell$-th service antenna, its LoS probability in \eqref{eqn05} is given by 
\begin{equation}\label{eqn08}
p=\mathcal{P}_{\textsc{los}}(d^\textsc{2d}).
\end{equation}
Moreover, the probability for two service-antennas to have the same NLoS/LoS state is also specified in \cite[(7.4-5)]{3gpp.38.901}, i.e.,
\begin{equation}\label{eqn09}
	\mathcal{P}(\ell, m) = \exp\left(-\frac{\Delta(\ell, m)}{d_\mathrm{cor}}\right).
\end{equation}
When the ELAA-mMIMO is a ULA antenna system, \eqref{eqn09} becomes
\begin{equation}\label{eqn10}
	\mathcal{P}(\ell, m) = \exp\left(-\frac{\lambda|\ell-m|}{2d_\mathrm{cor}}\right),
\end{equation}
where $\lambda$ is the carrier wavelength. 
Using \eqref{eqn05} and \eqref{eqn09}, we can compute the LoS probability $p_m$ for the $m$-th antenna as
\begin{equation}\label{eqn11}
p_m=p\mathcal{P}(\ell, m)+(1-p)(1-\mathcal{P}(\ell, m)).
\end{equation}
Plugging \eqref{eqn09} into \eqref{eqn11} with some tidy-up work, we can obtain the result \eqref{eqn06}. 

The result \eqref{eqn06} has rich physical implications:

{\em 1)} For the case of $p>0.5$, the $\ell$-th service antenna is more likely on the LoS state. 
Since $(2p-1)>0$, the probability for the $m$-th service antenna to be on the LoS state decreases exponentially with the distance $\Delta(\ell, m)$.
This forms the exponentially decaying window for the NLoS/LoS spatial correlation. 

{\em 2)} For the case of $p<0.5$, the $\ell$-th service antenna is less likely on the LoS state. 
Given $(2p-1)<0$, the probability for the $m$-th service antenna to be on the LoS state increases exponentially with the distance $\Delta(\ell, m)$.
Again, this implies the exponentially decaying window for the NLoS/LoS spatial correlation. 

{\em 3)} For the case of $p=0.5$, the distance $\Delta(\ell, m)$ has no impact on the NLoS/LoS correlation. 
Each link has $50\%$ probability for NLoS and $50\%$ for LoS. 
This is the case we should (and can) avoid in the channel realization (see the pseudocode in this section).

\begin{cor}\label{cor1}
Suppose that the binary random variable $b_m$ obeys the correlated Bernoulli distribution specified in {\em Proposition \ref{prop1}}. 
For the ULA, the size of NLoS/LoS window ($L$) obeys the following random distribution
\begin{equation}\label{eqn12}
f(L)= \exp\left(\dfrac{-\lambda(L^2 -L)}{4d_{\mathrm{cor}}}\right) - \exp\left(\dfrac{-\lambda(L^2 + L)}{4d_{\mathrm{cor}}}\right).
\end{equation}
\end{cor}
\begin{IEEEproof}
The proof of \eqref{eqn12} can be based on the computing of the probability for $L$ service antennas to share the same NLoS/LoS state. 
The mathematical derivations are rather trivial. They are abbreviated here for the sake of space limit. 
\end{IEEEproof}

\subsection{Normalization of Non-Stationary Channels}
In link-level computer simulations, wireless channels must be appropriately normalized for the sake of fair comparisons. 
The normalization of WSSUS channels is often implemented by
\begin{equation}\label{eqn13}
	\overline{\mathbf{H}} = \frac{\mathbf{H}}{\sqrt{\mathbb{E}\Vert\mathbf{H}\Vert^2}},
\end{equation}
where $\mathbb{E}(\cdot)$ stands for the expectation, $\|\cdot\|$ for the Frobenius norm, and $\overline{\mathbf{H}}$ the normalized channel transition matrix. 
For WSSUS channels, the term $\mathbb{E}\|\mathbf{H}\|^2$ can be simply computed as
\begin{equation}\label{eqn14}
\mathbb{E}\|\mathbf{H}\|^2 = \sum_{m=0}^{M-1}\sum_{n=0}^{N-1}\mathbb{E}\vert h_{m,n}\vert^2,
\end{equation}
where $h_{m,n}$ is the $(m,n)$-th element of $\mathbf{H}$. For identically distributed channels, we have $\mathbb{E}| h_{m,n}|^2=\sigma_h^2$ (a constant), and thereby having $\mathbb{E}\|\mathbf{H}\|^2 =MN\sigma_h^2$.
Unfortunately, this is not exactly the case for randomly realized non-stationary channels. 

Due to the channel spatial non-stationarity, every element in $\mathbf{H}$ is a non-stationary process in their Monte-Carlo trials. 
They can be NLoS in one trial and LoS in another; as described in \eqref{eqn04}. 
With the model \eqref{eqn04}, the power of each element can be computed by
\begin{IEEEeqnarray}{ll}
\mathbb{E}|h_{m}|^2 &=\mathbb{E}(b_{m})\mathbb{E}(\varepsilon_{m}^\textsc{los})\mathbb{E}|h_{m}^\textsc{los}|^2 +\nonumber\\
&\quad (1-\mathbb{E}(b_{m})) \mathbb{E}(\varepsilon_{m}^\textsc{nlos})\mathbb{E}|h_{m}^\textsc{nlos}|^2.\label{eqn016}
\end{IEEEeqnarray}
With \eqref{eqn02} and \eqref{eqn03}, we can obtain
\begin{equation}\label{eqn017}
\mathbb{E}|h_{m}^\textsc{los}|^2=\frac{\alpha^2}{d_m^{2q}},~ \mathbb{E}|h_{m}^\textsc{nlos}|^2=\frac{\beta^2}{d_m^{2\rho}}.
\end{equation}
Moreover, $b_m$ obeys the Bernoulli distribution, and thus we have 
\begin{equation}\label{eqn018}
\mathbb{E}(b_m)=p_m,
\end{equation}
where $p_m$ can be obtained from the 3GPP document with respect to various wireless environments such as something related to \eqref{eqn07}. 
Finally, the shadowing-related terms, i.e., $\mathbb{E}(\varepsilon_{m}^\textsc{los})=\exp(\sigma_\textsc{los}^2/2)$ and $\mathbb{E}(\varepsilon_{m}^\textsc{nlos})=\exp(\sigma_\textsc{nlos}^2/2)$, can be computed from the log-normal distribution specified right after \eqref{eqn04}.

\subsection{Pseudocode for the Implementation}
Now, we reach the stage to implement the proposed channel model in computer simulations. 
The following pseudocode is used to describe how to randomly generate the channel vector $\mathbf{h}$ in a Monte Carlo trial.  

\begin{algorithm}[ph]
{\footnotesize
	\renewcommand{\thealgorithm}{}
	\caption{{Random Realization of} $\mathbf{h}$} 
	\begin{algorithmic}[1]\label{alg01}
		\renewcommand{\algorithmicrequire}{\textbf{Input:}} 
		\REQUIRE~\\ $M$: the number of service antennas;\\ $d_\ell^{\textsc{2d}}$: the 2-D distance used in \eqref{eqn07};\\ 
		$\lambda, d_\mathrm{cor}$: parameters used in \eqref{eqn10};
		\renewcommand{\algorithmicrequire}{\textbf{Output:}} 
		\REQUIRE~\\$\mathbf{h}$: the channel vector;
		\renewcommand{\algorithmicensure}{\textbf{START}}
		\ENSURE  
		\STATE {\bf let} $\ell=0$; call \eqref{eqn07} to compute $\mathcal{P}_{\textsc{los}}(d_\ell^{\textsc{2d}})$;
		\STATE {\bf let} $p=\mathcal{P}_{\textsc{los}}(d_\ell^{\textsc{2d}})$ and generate $b_\ell$ according to the Bernoulli distribution in \eqref{eqn05}; {\bf let} $m=\ell+1$;
		\STATE {Generate} $b_m$ according to the distribution in \eqref{eqn10};
		\STATE {\bf if} $b_m=b_\ell$, {\bf then} $m\leftarrow m+1$; {\bf otherwise} $\ell\leftarrow m$, {\bf goto} step 2; 
		\STATE {\bf repeat} step 3 until $m=M$; 
		\STATE {Generate} $\mathbf{h}$ using \eqref{eqn04} and conduct the normalization;
		\renewcommand{\algorithmicensure}{\textbf{END}}
		\ENSURE
	\end{algorithmic} 
	}
\end{algorithm}

It is worthwhile to highlight the step 4. When $b_m\neq b_\ell$, it means that $b_m$ is uncorrelated with $b_\ell$. 
Hence, the realization of $b_m$ conditioned on $b_\ell$ is no longer reliable,  and we must independently regenerate $b_m$ by going back to step 2. 

\subsubsection*{Remark 1}
The algorithm described in the pseudocode is robust to the case of $\mathcal{P}_{\textsc{los}}(d_\ell^{\textsc{2d}})=0.5$. 
This is because $b_\ell$ is still generated according to the equal-probability, and $b_m$ is only dependent on \eqref{eqn10}. 

\subsubsection*{Remark 2} 
In practice, the NLoS/LoS state is determined by the wireless environment rather than the number of service antennas. 
For instance there is a window having $M_w$ antennas. The length of the window is $\Big(\frac{M_w-1}{2}\Big)\lambda$ meter. 
In this case, even if the antenna spacing is not $\lambda/2$ or the antenna array is not an ULA, the length of the window can also be generated according to \eqref{eqn12}. With the window length, no matter how many antennas in the window, they should have the same NLoS/LoS state. 

\subsubsection*{Remark 3} 
This channel model is also suitable for ELAA-mMIMO with a large rectangular antenna array (URA); as described in \cite{BJORNSON20193}. 
The window can be generated from the lowest row of the URA. 
For those antennas located in the same column, if the lowest antenna is in the LoS state, then others in the column are also in the LoS state. 
If a lower antenna is in the NLoS state, then the antenna at one row above can be generated according to the Bernoulli distribution \eqref{eqn05}. 

Finally, our channel model in its current form does not incorporate antenna correlations in small-scale fading. 
This is an interesting work that will be studied in our future work. 

\begin{figure}[t]
	\centering
	\subfigure{
		\begin{minipage}[t]{0.23\textwidth}	
			\label{fig02a}
			\centering
			\includegraphics[width=4.75cm]{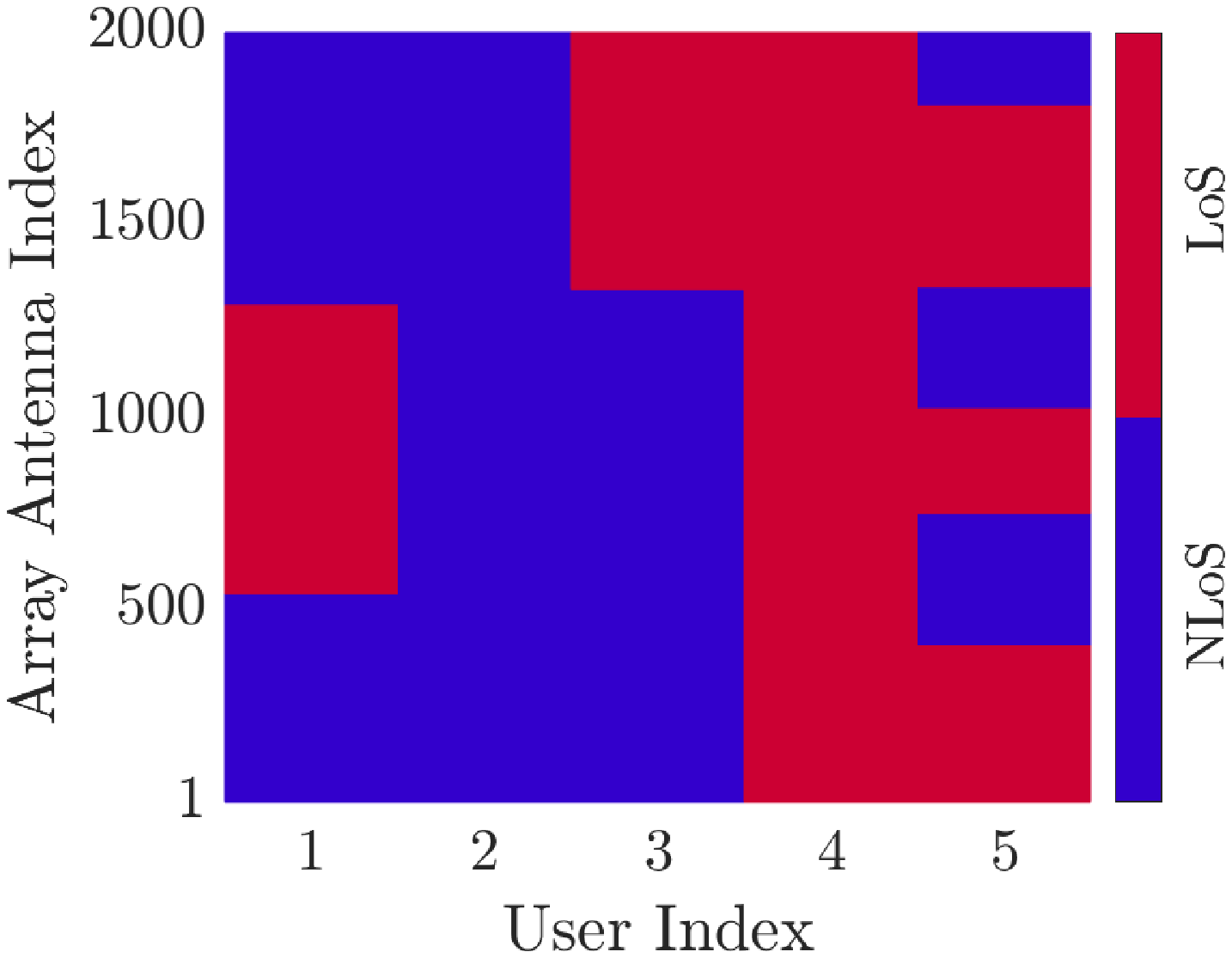}	
	\end{minipage}}
	\subfigure{
		\begin{minipage}[t]{0.232\textwidth}
			\label{fig02b}
			\centering
			\includegraphics[width=4.42cm]{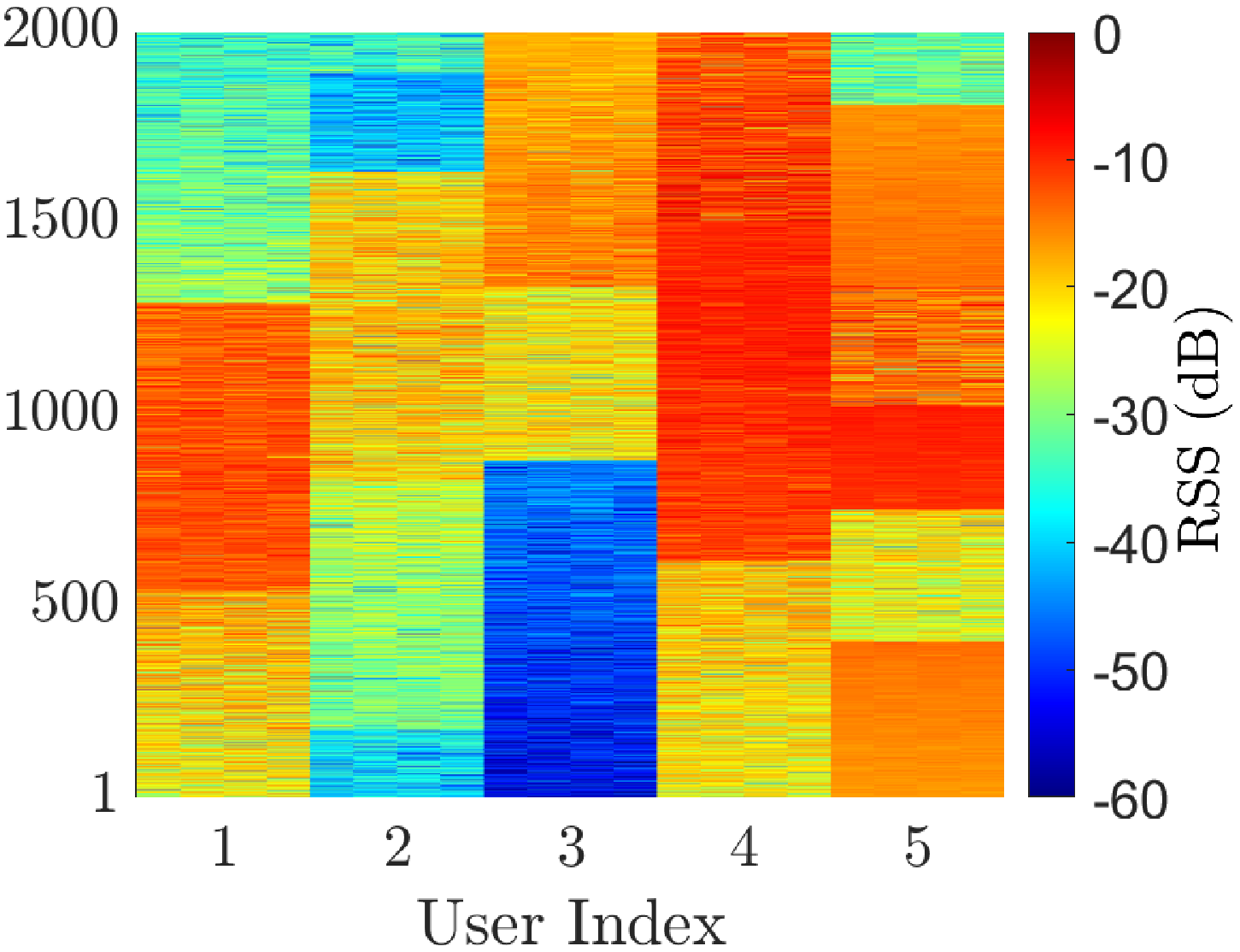}				   	
	\end{minipage}}    	
	\caption{\label{fig02} NLoS/LoS state and RSS from a user to ULA antennas. {\bf Left}: the NLoS/LoS state of every link. 
	{\bf Right}: the RSS (normalized by the transmit power) heat map.}
	\vspace{-2em}
\end{figure}

\section{Numerical Results and Discussion} \label{sec04}
In this section, the objectives are: {\em 1)} to showcase the proposed channel model through computer emulations, 
and {\em 2)} to study the statistical behavior of ELAA-mMIMO channel capacity using the proposed channel model.
In our case study, the ELAA-mMIMO is a large ULA depicted in \figref{fig01}, 
which has $M=2,000$ service antennas and the length of $85$ meters (on the $3.5$ GHz central frequency). 
The propagation environment is the UMi-street canyon. 
According to the \gls{3gpp} document \cite{3gpp.38.901}, the system parameters are configured by: $d_{\text{cor}}=5,000$, 
$\mu_\kappa = 2.07$, $\sigma_\kappa = 1.15$, $\sigma_\textsc{nlos} = 1.80$, $\sigma_\textsc{los} = 0.92$, $\overline{d_\mathrm{1}} = 18 $, 
$\overline{d_\mathrm{2}} = 36$, $\alpha = 0.020$, $\beta = 0.007$, $\rho = 1.765$, and $q = 1.050$. 
The objective sets two items for our numerical study. 

\subsubsection*{Study Item 1}
This study item aims to showcase the proposed channel model using the MATLAB ``siteviewer'' emulation toolbox.  
We import the 3-D map of the University of Surrey, Stag Hill Campus, into the toolbox and choose the avenue similar to the UMi-street canyon scenario (see \figref{fig01}).
The height of ULA is set to $10$ meters. 
For users located at different places on the map, they will experience different NLoS/LoS non-stationarity. 
We use the proposed channel model to generate the non-stationary channel in this environment and illustrate in \figref{fig02} the NLoS/LoS state and RSS (normalized by the transmit power) of every user-to-service antenna link when users are located at different locations on the map.

\begin{figure*}[t]
	\centering
	\subfigure{
		\begin{minipage}[t]{0.49\textwidth}	
			\label{fig03a}
			\centering
			\includegraphics[width=8.5cm]{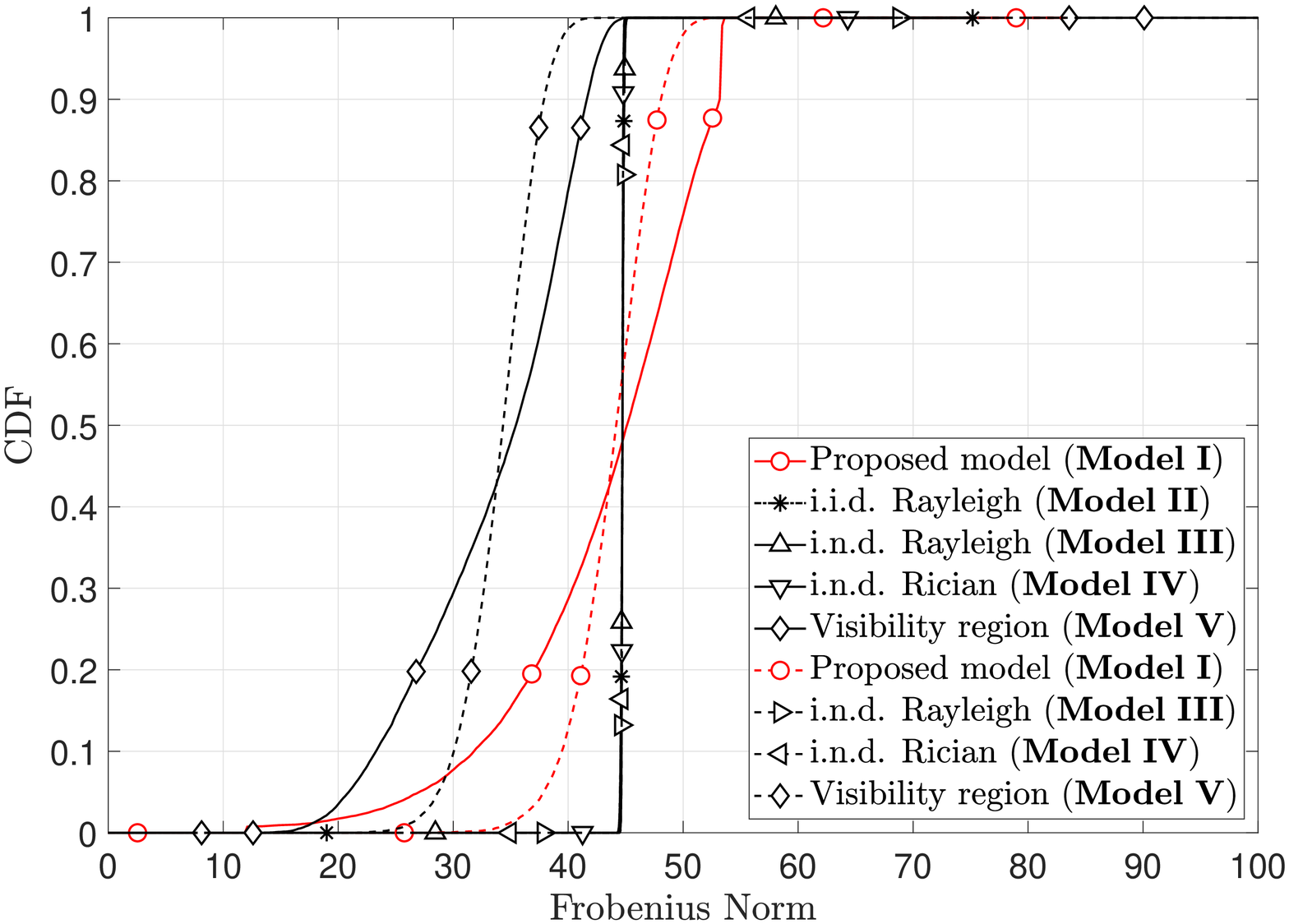}	
	\end{minipage}}
	\subfigure{
		\begin{minipage}[t]{0.49\textwidth}
			\label{fig03c}
			\centering
			\includegraphics[width=8.5cm]{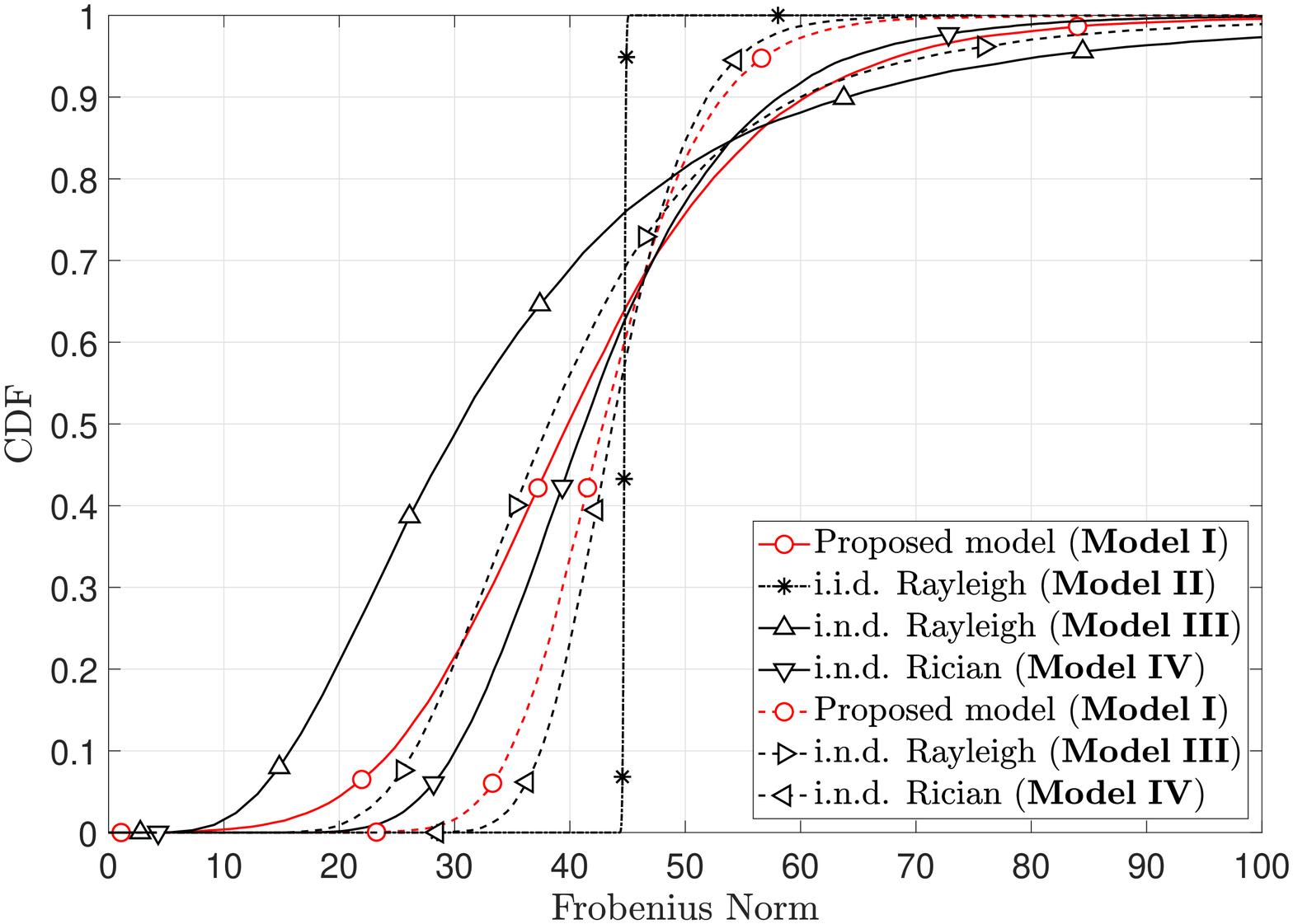}			   	
	\end{minipage}}	
	\caption{\label{fig03} The \gls{cdf} of Frobenius norm for different channel models. {\bf Left:} no shadowing, {\bf Right:} with shadowing.
		\protect\tikz[baseline]{\protect\draw[line width=0.2mm] (0,.5ex)--++(0.6,0) ;}~high density;
		\protect\tikz[baseline]{\protect\draw[line width=0.2mm, dashed] (0,.5ex)--++(0.6,0) ;}~low density.}
		\vspace{-1em}
\end{figure*}

\begin{figure*}[t]
	\centering
	\subfigure{
		\begin{minipage}[t]{0.49\textwidth}	
			\label{fig04a}
			\centering
			\includegraphics[width=8.5cm]{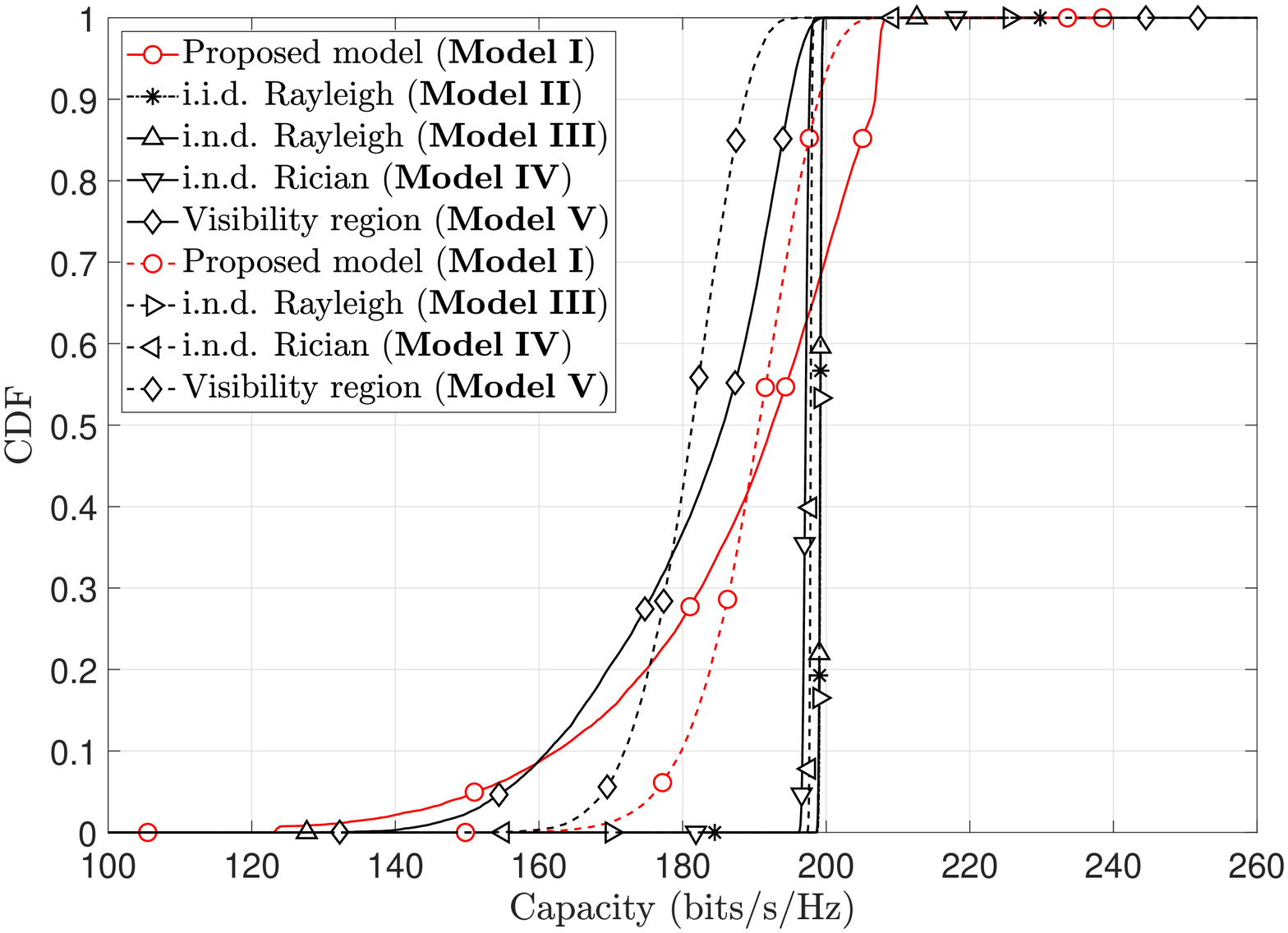}	
	\end{minipage}}
	\subfigure{
		\begin{minipage}[t]{0.49\textwidth}
			\label{fig04b}
			\centering
			\includegraphics[width=8.5cm]{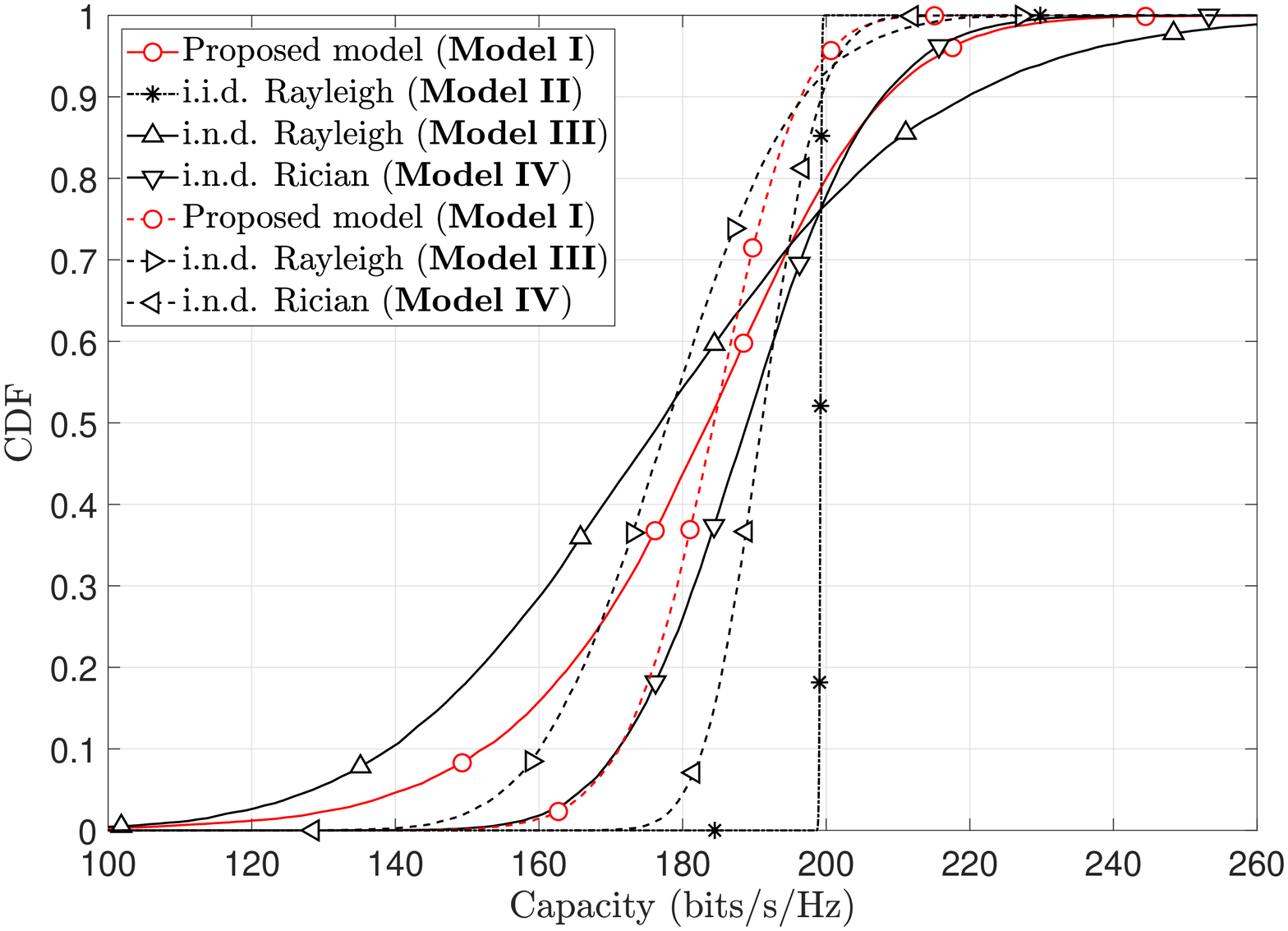}			   	
	\end{minipage}}		
	\caption{\label{fig04} The \gls{cdf} of channel capacity at the average-\gls{snr}$=10$ dB. {\bf Left:} no shadowing, {\bf Right:} with shadowing.
		\protect\tikz[baseline]{\protect\draw[line width=0.2mm] (0,.5ex)--++(0.6,0) ;}~high density;
		\protect\tikz[baseline]{\protect\draw[line width=0.2mm, dashed] (0,.5ex)--++(0.6,0) ;}~low density.}
		\vspace{-1em}
\end{figure*}

Basically, the randomly generated channel can well reflect the NLoS/LoS non-stationarity.
For instance on the user index 1, this is corresponding to the user place shown in \figref{fig01}, where all LoS links are in the middle of the ULA. 
For a user located at somewhere between the square building and the ULA, all links are in the LoS state (corresponding to the user index 4 in \figref{fig02}). 
For a user located at somewhere right behind a building, all links are in the NLoS state (corresponding to the user index 2 in \figref{fig02}). 
In addition, the RSS heat map shows that most of the signal power is distributed on those LoS links. 
Some of NLoS links even experience deep fades, and they form an invisible region (see user index 3). 

\subsubsection*{Study Item 2}
This study item aims to investigate the statistical behavior of ELAA-mMIMO channel capacity. 
The following channel models are considered in our study:
\subsubsection*{{\bf Model I}}
our proposed non-stationary channel model;
\subsubsection*{{\bf Model II}}
i.i.d. Rayleigh fading channel. This is a special case of  {\bf Model I} by setting $\mathcal{P}_\textsc{los}(d^\mathrm{2D}) = 0$, $d_m=1, _{\forall m}$, $\alpha = 1$, $\sigma_\textsc{nlos} = 0$;
\subsubsection*{{\bf Model III}}
the non-stationary Rayleigh fading channel in \cite{8644126}. This is a special case of {\bf Model I} by setting $\mathcal{P}_\textsc{los}(d^\mathrm{2D}) = 0$ and $\sigma_\textsc{nlos} = 0$;
\subsubsection*{{\bf Model IV}}
the non-stationary Rician fading channel \eqref{eqn03}. This is also a special case of {\bf Model I};
\subsubsection*{{\bf Model V}}
the single-cluster channel model proposed in \cite{9145378}. The visibility region is randomly distributed on the ULA with its length following the log-normal distribution ($L\sim \mathcal{LN}(\ln4, \ln0.2)$ in our study).

Note that it is not our aim here to compare different channel models in terms of their accuracy or practicality. 
The accuracy and practicality of our model have been well reflected in \textit{Study Item 1}. 
Instead, we are interested in the cumulative distribution function (CDF) of channel capacity and the channel matrix-norm in different channel models. 
To this purpose, we set $K=5$ users with each having $4$ antennas for signal transmission (i.e., $N=4$, $K=20$). 
Antennas at the same user are assumed to have the same NLoS/LoS states as well as shadowing effects. 
Moreover, users are placed in front of the ULA, and they are uniformly distributed on a line that is parallel to the ULA. 

There are two cases of user density in our study. One is the high density case where all users are located within a $1$-meter range, 
and the other is the low density case, where the range is $20$-meter long. 
In the high density case, users share the same NLoS/LoS state for all of their links; 
in the low-density case, the NLoS/LoS states for different users are independently generated.

\figref{fig03} shows the \gls{cdf} of the Frobenius norm of the normalized channel matrix $\overline{\mathbf{H}}$ in \eqref{eqn13}. 
Without considering the shadowing, channel hardening effects can be observed in {\bf Model II} (WSSUS), {\bf Model III}, and {\bf Model IV}.  
This is because {\bf Model III} and {\bf Model IV} have their non-stationarity mainly coming from the spherical-wave propagation. 
For fixed user locations, their large-scale fading components are temporally stationary, and their small-scale fading is stationary as well. 
In contrast, the channel hardening effect is much weaker in the proposed channel model and {\bf Model V}.  
This attributes to the presence of mixed NLoS/LoS states in our model and the spatial visibility region in {\bf Model V}. 
Due to the same reason, the presence of non-stationary shadowing could also largely weaken the channel hardening effect 
\footnote{The shadowing effect was not considered in {\bf Model V} since it was not a part of the original model.}. 
When comparing the cases with different user density, it is found that user separation helps the channel hardening in non-stationary channels.
This is because users are only partially connected in the low-density case, and each user enjoys more spatial diversity. 

Given that the concept of channel hardening was originally defined on the channel capacity,  we plot in \figref{fig04} the \gls{cdf} of channel capacity.
For all exhibited results, the average-\gls{snr} is set to $10$ dB. 
It can be found that the statistical behavior of channel capacity well coincides with the statistical behavior of Frobenius norm. 
In non-stationary channels, the low-density case shows higher channel capacity than the high-density case. 
Again, this attributes to the sparsity of non-stationary channels, which must be carefully considered in the ELAA-mMIMO transceiver design.  

\section{Conclusion and Outlook}\label{sec05}
In this paper, a novel non-stationary channel model has been proposed and evaluated for ELAA-mMIMO. 
Unlike current non-stationary channel models, the proposed model allows a mix of NLoS/LoS links between the user and service antennas. 
The correlation of NLoS/LoS state between links is modeled as an exponentially decaying window. 
Computer emulations, based on a 3-D real-life map, have shown that the proposed channel model can capture almost all spatially non-stationary properties, including spherical wavefront, shadowing effects, and many physical differences between NLoS and LoS states. 
A pseudocode has been provided to demonstrate the channel realization in Monte-Carlo simulations. 
It is simple and easy to implement on the link level or multi-link level. 
Moreover, computer simulations have been conducted to evaluate the statistical behavior of ELAA-mMIMO channel capacity (and Frobenius norm) in non-stationary channels. 
It has been shown that the channel non-stationarity is detrimental to the mMIMO channel hardening. 
Nevertheless, user separation plays a constructive role to the channel capacity taking the advantage of the channel spatial sparsity. 

Due to the page limit, we are unable to present some other interesting issues about non-stationary channels.  
Those include multiuser spatial-correlations, antenna correlations in small-scale fading, eigenvalue analysis, non-stationary channel sparsity, 
as well as linear transceiver design. All of these will be presented in the journal version of this work. 

\section*{Acknowledgement}
This work is partially funded by the 5G Innovation Centre and 6G Innovation Centre.

\ifCLASSOPTIONcaptionsoff
\newpage
\fi

\bibliographystyle{IEEEtran}
\bibliography{../mMIMO}		
\end{document}